\documentclass[12pt]{article}

\usepackage{axodraw,amsmath,amssymb,graphicx}

\newcommand{\beq}{\begin{equation}}
\newcommand{\eeq}{\end{equation}}
\newcommand{\bea}{\begin{eqnarray}}
\newcommand{\eea}{\end{eqnarray}}
\newcommand{\ba}{\begin{array}}
\newcommand{\ea}{\end{array}}
\newcommand{\bec}{\begin{center}}
\newcommand{\eec}{\end{center}}
\newcommand{\bei}{\begin{itemize}}
\newcommand{\eei}{\end{itemize}}

\newcommand{\Fig}[1]{Fig.~\ref{fig:#1}}

\begin{document}

\begin{flushright}
SACLAY-T08/070\\
SISSA-08/2008/EP\\
\end{flushright}
\vspace{0.5cm}

\begin{center}
{\Large	\bf
A new, direct link between the baryon}\\ 
\vspace{0.3cm}
{\Large  \bf asymmetry and neutrino masses}\\
\vspace{1.5cm}

{\bf Michele Frigerio$^a$, Pierre Hosteins$^b$, St\'ephane Lavignac$^a$ \\
and Andrea Romanino$^c$}\\
\vspace{0.3cm}
\end{center}

{\sl $^a$ Institut de Physique Th\'{e}orique, CEA-Saclay, 
91191 Gif-sur-Yvette Cedex, France~\footnote{Laboratoire de la Direction
des Sciences de la Mati\`ere du Commissariat \`a l'Energie Atomique
et Unit\'e de Recherche Associ\'ee au CNRS (URA 2306).}}\\
\indent {\sl $^b$ Department of Physics, University of  Patras,
GR-26500 Patras, Greece}\\
\indent {\sl $^c$ International School for Advanced Studies (SISSA)
and INFN, I-34013 Trieste, Italy}\\
\vspace{0.5cm}

\abstract{
We point out that, in a class of $SO(10)$ models with matter fields
in $\bf 16$ and $\bf 10$ representations
and type II realization of the seesaw mechanism,
the light neutrino masses and the CP asymmetry needed for leptogenesis
are controlled by one and the same set of couplings.
The generated baryon asymmetry then directly depends
on the low-energy neutrino parameters, with no unknown
seesaw-scale flavour parameters involved; in particular,
the necessary CP violation is provided by the CP-violating
phases of the lepton mixing matrix.
We compute the CP asymmetry in triplet decays for this scenario
and show that it can lead to successful leptogenesis.
\
}

\section{Introduction}

The seesaw mechanism~\cite{seesaw} nicely connects
two pieces of observation that remain unexplained
in the Standard Model: the smallness of neutrino masses
and the fact that our universe contains almost no antimatter.
Indeed, the same heavy states that are responsible for the
generation of small Majorana masses for the Standard Model
neutrinos in the seesaw mechanism can be the source
of the observed baryon asymmetry through their decays. 
The lepton asymmetry generated
in the out-of-equilibrium decays of the heavy states
is then partially converted into a baryon asymmetry by the
non-perturbative sphaleron processes~\cite{KRS85}.
This mechanism is known as baryogenesis via
leptogenesis~\cite{FY86}.
The role of the heavy states is generally played
by right-handed neutrinos (type I seesaw mechanism) and/or by scalar
$SU(2)_L$ triplets (type II seesaw mechanism~\cite{typeII}).

In spite of this intimate connection between neutrino masses
and leptogenesis, there is little correlation, in general, between
the generated baryon asymmetry and the neutrino parameters
that can be measured at low energy. For instance,
in the Standard Model augmented with a type I seesaw mechanism,
the CP asymmetry in right-handed neutrino decays is not directly related
to the CP-violating phases of the lepton mixing (PMNS)
matrix~\cite{correlations}.
Even in the constrained framework of $SO(10)$ Grand Unified Theories
(GUTs)~\cite{SO10}, there is generally enough freedom
to spoil any relationship of this kind.
Exceptions to this general statement are possible at the price
of strong assumptions on the seesaw parameters~\cite{assumptions}.

In this paper, we present a new leptogenesis scenario in which
the generated baryon asymmetry is directly related to the low-energy
neutrino parameters, with no dependence on unknown seesaw-scale
flavour parameters. In particular, the relevant CP asymmetry
depends on the CP-violating phases of the PMNS matrix.
This scenario is realized in $SO(10)$ models
with Standard Model fermions
split among $\bf 16$ and $\bf 10$ representations
and type II realization of the seesaw mechanism.

The paper is organized as follows. In Section~\ref{sec:model},
we describe the class of $SO(10)$ models in which our leptogenesis
scenario takes place. In Section~\ref{sec:leptomec}, we present
this scenario and compute the relevant CP asymmetry.
We then discuss its dependence on the light neutrino parameters.
In Section~\ref{sec:efficiency}, we write the Boltzmann equations
and estimate the efficiency factor that determines the final baryon asymmetry.
We also comment on lepton flavour effects and on the contributions of other
heavy states present in the model. Finally, we present our conclusions
in Section~\ref{sec:conclusions}.

\section{The $SO(10)$ framework \label{sec:model}}

In standard $SO(10)$ unification, all Standard Model fermions of a given
generation reside in a $\bf 16$ representation, together with a right-handed
neutrino. The charged fermion and Dirac neutrino mass matrices receive
contributions from Yukawa couplings of the form ${\bf 16}_i {\bf 16}_j {\bf H}$
(where ${\bf H} = {\bf 10}, {\bf \overline{126}}$ and/or ${\bf 120}$)
as well as from non-renormalizable operators suppressed by powers
of a scale $\Lambda \gg M_{GUT}$.
Majorana masses for the right-handed neutrinos
are generated either from ${\bf 16}_i {\bf 16}_j {\bf \overline{126}}$,
or from the non-renormalizable operators
${\bf 16}_i {\bf 16}_j {\bf\overline{16}}\, {\bf\overline{16}} / \Lambda$.
This leads to small Majorana masses for the light neutrinos via the type I
seesaw mechanism. In addition, light neutrino masses may also receive
a type II seesaw contribution from the exchange of
the electroweak triplet contained in the $\bf \overline{126}$ representation.
Leptogenesis in this framework has been studied by many
authors~\cite{le10}, both in the type I and in the type I+II cases. In spite
of the existence of constraining $SO(10)$ mass relations, the predictivity
of these models for leptogenesis is generally limited, due in particular
to the presence of physical high energy phases contributing
to the CP asymmetry.

In this paper, we consider a different class of (supersymmetric)
$SO(10)$ models, in which the Standard Model fermions are split
among $\bf 16$ and $\bf 10$ matter multiplets~\cite{10matter,10MdMe}.
More precisely, we are interested in a subclass of these models
in which neutrino masses arise from a pure type II seesaw mechanism.
The advantage of these models is that, as we are going to see,
they lead to a more predictive leptogenesis.
To be specific, let us consider the following superpotential:
\bea
  W & = & \frac 12\, y_{ij} {\bf 16}_i {\bf 16}_j {\bf 10} + h_{ij} {\bf 16}_i {\bf 10}_j {\bf 16}
    + \frac 12\, f_{ij} {\bf 10}_i {\bf 10}_j {\bf 54}  \nonumber \\
  & & +\, \frac 12\, \sigma {\bf 10}\, {\bf 10}\, {\bf 54} + \frac 1 2\, M_{54} {\bf 54}^2 +\, \cdots\ ,
\label{supmin}
\eea
where ${\bf 16}_i$ and ${\bf 10}_i$ ($i=1,2,3$) are matter representations,
${\bf 10}$, ${\bf 16}$ and ${\bf 54}$ are Higgs representations,
and we imposed a matter parity to restrict the form of the superpotential.
The splitting of the matter representations
${\bf 16}_i \equiv ({\bf 10}^{16}, {\bf \overline 5}^{16}, {\bf 1}^{16})_i$ and
${\bf 10}_i \equiv ({\bf \overline 5}^{10}, {\bf 5}^{10})_i$ is realized by the
vev $v^{16}_1$ of the $SU(5)$-singlet component of $\bf 16$, which gives
large $SU(5)$-invariant masses $M_i = h_i v^{16}_1$ to the
(${\bf \overline 5}^{16}_i$, ${\bf 5}^{10}_i$)  pairs.
In the following, we shall refer to the components of ${\bf 5}^{10}_i$
as heavy antilepton doublets $L^c_i$ (with lepton number $L = - 1$)
and heavy quark singlets $D_i$ (with baryon number $B = + 1/3$), respectively:
\beq
  {\bf 5}^{10}_i\ \equiv\ (L^c, D)_i\ .
\eeq

The Standard Model fermions are identified as:
\beq
  {\bf 10}^{16}_i\ =\ (Q, u^c, e^c)_i\ , \qquad
  {\bf \overline 5}^{10}_i\ =\ (L, d^c)_i\ ,
\label{eq:SM_fermions}
\eeq
(the remaining ${\bf 16}_i$ component being the right-handed neutrino:
${\bf 1}^{16}_i\ =\ \nu^c_i$), and their mass matrices are given by,
at the renormalizable level:
\beq
  M_u\ =\ y\, v^{10}_u\ , \qquad M_d\ =\ M^T_e\ =\ h\, v^{16}_d\ ,
\label{eq:M_ude}
\eeq
where $v^{10}_u$ and $v^{16}_d$ are electroweak-scale vevs
associated with the $SU(2)_L$ doublet components $H^{10}_u$
and $H^{16}_d$ of $\bf 10$ and $\bf 16$, respectively\footnote{We
assume here that the light Higgs doublets $H_u$ and $H_d$
contain a significant component of $H_u^{10}$ and $H_d^{16}$,
respectively. We checked that this naturally arises in simple 
realizations of the doublet-triplet splitting.}. As is well known,
the GUT-scale mass relation $M_d = M^T_e$ is not consistent
with the measured values of the charged lepton and down quark
masses, and must be corrected by non-renormalizable operators.

As for neutrino masses, they arise from a pure type II seesaw mechanism,
in contrast with standard $SO(10)$ unification where the type I seesaw
contribution is always present.
Indeed, Eqs.~(\ref{supmin}) and (\ref{eq:SM_fermions}) imply that
the Dirac mass matrix vanishes at the renormalizable level,
while the $\bf 54$ Higgs multiplet contains a pair of electroweak triplets
($\Delta$, $\overline \Delta$) with the requisite couplings
$\frac 1 2 f_{ij} L_i L_j \Delta$ and $\frac 1 2 \sigma H^{10}_u H^{10}_u \overline \Delta$
to mediate a Majorana mass term for the Standard Model neutrinos:
\beq
  m_\nu\ =\ \frac{\sigma (v_u^{10})^2}{2M_\Delta}\, f\ ,
\label{mnu}
\eeq
with $M_\Delta = M_{54}$. In this paper, we neglect possible subdominant
type I contributions to Eq.~(\ref{mnu}) coming from non-renormalizable
operators.
Note that this realization of the type II seesaw
mechanism is very different from standard $SO(10)$ unification with
a $\bf 126 + \overline{126}$ pair.

The above conclusions hold as long as the ${\bf \overline 5}^{10}_i$'s
do not mix with the ${\bf \overline 5}^{16}_i$'s.
This assumes in particular that the ${\bf 54}$ Higgs multiplet in
Eq.~(\ref{supmin}) does not acquire a GUT-scale vev,
and that direct mass terms $M_{ij} {\bf 10}_i {\bf 10}_j$ are forbidden
(which can easily be achieved by a global symmetry).
In the opposite case, the light $L_i$ and $d^c_i$
are combinations of the corresponding components of
${\bf \overline 5}^{10}_i$ and ${\bf \overline 5}^{16}_i$, and many of
the features described above are modified. In particular, the  GUT-scale
relation $M_d = M^T_e$ no longer hold~\cite{10MdMe}, and neutrino
masses receive both type I and type II seesaw contributions.
Since we are interested in the more predictive type II case,
we shall not consider this possibility any longer.

Let us close this section with some remarks about the seesaw scale.
The observed frequency of atmospheric neutrino oscillations indicates
that $M_\Delta$ in Eq.~(\ref{mnu}) should lie significantly below the GUT
scale, $M_\Delta \lesssim 5 \times 10^{14}$ GeV. On the other hand,
it is difficult to maintain gauge coupling unification with incomplete
$SU(5)$ representations at an intermediate scale.
Since ${\bf 54} = {\bf 24} + {\bf 15} + {\bf \overline{15}}$ under $SU(5)$,
with $\Delta \subset {\bf 15}$ ($\overline \Delta \subset {\bf \overline{15}}$),
this suggests two possibilities\footnote{Note that no component
of the ${\bf 54}$ supermultiplet can mediate proton decay, so an
intermediate-scale ${\bf 54}$ or ${\bf 15} + {\bf \overline{15}}$ pair
does not cause any phenomenological problem.}:
(i) the whole $\bf 54$ representation
lies below the GUT scale ($M_\Delta = M_{54} \ll M_{GUT}$);
(ii) the ${\bf 24}^{54}$ component of $\bf 54$ is split from the
(${\bf 15}^{54}$, ${\bf \overline{15}}^{54}$) pair
($M_\Delta = M_{15} \ll M_{24}$). The first possibility can be realized
by generating an effective mass for the $\bf 54$ from the
non-renormalizable operator ${\bf \overline{16}\, 16\, 54\, 54} / \Lambda$,
while forbidding a direct mass term
by an appropriate symmetry. The second possibility
requires a splitting between $M_{24}$ and $M_{15}$. This can be
achieved e.g. by an additional coupling $\bf 54\, 45\, 45$
involving a $\bf 45$ Higgs representation with a GUT-scale vev
along its $SU(5)$-singlet component, which will give
a large mass to the ${\bf 24}^{54}$ component of $\bf 54$.

Notice that the requirement of perturbative gauge coupling unification
constrains the values of $M_{15}$ and $M_{24}$.
Indeed, while the unification of gauge couplings
at $M_{GUT} \simeq 2 \times 10^{16}$ GeV is not spoiled
by the addition of complete $SU(5)$ representations to the
MSSM spectrum, their contribution to the beta function
coefficients increases the value of the unified gauge coupling.
Taking e.g. $M_{24} =10 M_{15} = 100 M_1$ (and noting that
Eq.~(\ref{supmin}) implies $M_i \propto m_{e_i}$, where the $M_i$
are the masses of the  (${\bf 5}^{10}_i$,${\bf \overline 5}^{16}_i$)
pairs), the requirement of perturbative unification, $\alpha_{GUT} < 1$,
sets a lower bound of $4 \times 10^{11}$ GeV on $M_{15}$.
This bound can be lowered if the
(${\bf 15}^{54}$, ${\bf \overline{15}}^{54}$) and/or ${\bf 24}^{54}$
multiplets are split, or if $M_{24} \gg 10 M_{15}$.

\section{The leptogenesis scenario \label{sec:leptomec}}

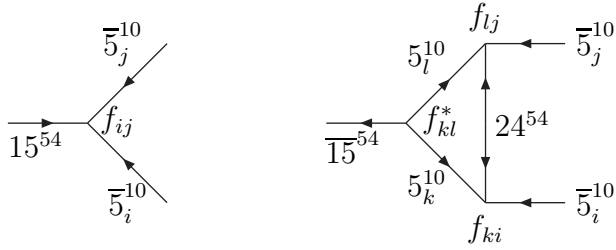
\begin{figure}
\begin{center}
\begin{picture}(130,80)(-60,10)
\ArrowLine(-120,50)(-90,50)
\ArrowLine(-60,80)(-90,50)
\ArrowLine(-60,20)(-90,50)
\Text(-110,42)[]{$15^{54}$}
\Text(-75,20)[]{$\overline{5}^{10}_i$}
\Text(-77,80)[]{$\overline{5}^{10}_j$}
\Text(-79,51)[]{$f_{ij}$}
\ArrowLine(30,50)(0,50)
\ArrowLine(90,80)(60,80)
\ArrowLine(90,20)(60,20)
\ArrowLine(60,50)(60,80)
\ArrowLine(60,50)(60,20)
\ArrowLine(30,50)(60,80)
\ArrowLine(30,50)(60,20)
\Text(10,41)[]{$\overline{15}^{54}$}
\Text(38,26)[]{$5^{10}_k$}
\Text(38,75)[]{$5^{10}_l$}
\Text(74,50)[]{$24^{54}$}
\Text(102,80)[]{$\overline{5}^{10}_j$}
\Text(102,20)[]{$\overline{5}^{10}_i$}
\Text(43,51)[]{$f^*_{kl}$}
\Text(60,90)[]{$f_{lj}$}
\Text(60,10)[]{$f_{ki}$}
\end{picture}
\end{center}
\caption{Supergraphs responsible (together with the similar supergraphs
obtained upon replacing ${\bf 15}^{54} \leftrightarrow {\bf \overline{15}}^{54}$
and ${\bf 5}^{10}_i \leftrightarrow {\bf \overline 5}^{10}_i$) for the CP
asymmetries in the decays of the components of the
(${\bf 15}^{54}$, ${\bf \overline{15}}^{54}$) supermultiplets.}
\label{fig1}\end{figure}

The class of $SO(10)$ models described in the previous section
contains several heavy states that can
contribute to leptogenesis through their decays.
Let us first consider the components of the
(${\bf 15}^{54}$, ${\bf \overline{15}}^{54}$) multiplets, which as discussed
above must be significantly lighter than $M_{GUT}$ (we shall address
the case of the ${\bf 24}^{54}$ multiplet in Section~\ref{other}).
Under $SU(3)_c \times SU(2)_L \times U(1)_Y$,
the $\bf 15$ representation decomposes as ${\bf 15} = (\Sigma, Z, \Delta)$,
where $\Sigma = (6,1)_{-2/3}$, $Z = (3,2)_{+1/6}$ and $\Delta = (1,3)_{+1}$.
Since ${\bf 10}_i {\bf 10}_j {\bf 54} \supset
{\bf \overline 5}^{10}_i {\bf \overline 5}^{10}_j {\bf 15}^{54}$,
each of these states can decay into a pair of light matter fields.
The supergraphs responsible for the asymmetries between these decays
and the CP-conjugated decays are represented in Fig.~\ref{fig1}, where, depending
on the decaying state, ${\bf 15}^{54}$, ${\bf \overline{15}}^{54}$,
${\bf 24}^{54}$, ${\bf 5}^{10}_i$ and ${\bf \overline 5}^{10}_i$ should
be replaced by their appropriate components.
Note that there is no self-energy contribution to the CP asymmetry,
since the interference of the self-energy diagram with the tree level
diagram is proportional to the real combination of couplings
$[{\rm Tr}(ff^*)]^2$.
If all three generations of ${\bf 5}^{10}_i$'s were massless
or degenerate in mass, the vertex contribution would be proportional
to ${\rm Im}[{\rm Tr}(ff^*ff^*)]=0$ and would therefore vanish as well.

However, this is not the case in our scenario, since the
(${\bf 5}^{10}_i$, ${\bf \overline 5}^{16}_i$) vector-like pairs
are heavy with strongly hierarchical masses.
Indeed, as follows from the second term in Eq.~(\ref{supmin}),
their masses are determined by the same couplings as the
charged lepton masses (up to possible corrections from
non-renormalizable operators modifying the mass relation $M_d = M^T_e$):
$M_i = h_i v_1^{16} \approx m_{e_i} (v_1^{16}/v_d^{16})$,
where the $h_i$ are the eigenvalues of the matrix of couplings
$h_{ij}$. Furthermore, the optical theorem tells us that only
the (${\bf 5}^{10}_i$, ${\bf 5}^{10}_j$) pairs with $M_{15}>M_i+M_j$
can contribute to the CP asymmetry. The vertex contribution is
therefore proportional to $\sum_{ij} c_{ij}\, \theta(M_{15} - M_i - M_j)\,
{\rm Im}[f_{ij}(f^*ff^*)_{ij}]$, where the coefficients $c_{ij}$ account
for the $M_i$-dependent loop function, and $\theta (x)$ is the
Heaviside function. Thus, thanks to the presence of heavy states
with hierarchical masses in the loop, the CP asymmetry does not vanish
in our scenario. This is very different from the usual triplet
leptogenesis scenarios~\cite{triplet_lepto}, in which two 
different sets of couplings are needed in order to obtain
a non-vanishing CP asymmetry.

Notice that (i) the three relevant $SU(5)$ couplings
${\bf 5}^{10}_i {\bf\overline{5}}^{10}_j {\bf 24}^{54}$, 
${\bf\overline{5}}^{10}_i {\bf\overline{5}}^{10}_j {\bf 15}^{54}$ and
${\bf 5}^{10}_i {\bf 5}^{10}_j {\bf \overline{15}}^{54}$ are determined by one
and the same matrix $f$;
(ii) the masses of the ${\bf 5}^{10}_i$'s are approximatively
known, up to an overall scale, in terms of charged lepton masses. 
These properties, which follow from the underlying $SO(10)$ structure,
make this model more predictive than usual leptogenesis scenarios.

\subsection{The CP asymmetry in triplet decays \label{epsico}}

In this section, we compute the CP asymmetry in triplet decays.
The relevant superpotential, inherited from the $SO(10)$
superpotential~(\ref{supmin}), reads:
\begin{eqnarray}
  W_{\Delta, \overline{\Delta}} & = & M_\Delta \mbox{Tr} (\Delta \overline{\Delta})
    + \frac 12\, f_{ij}\! \left( L^T_i i \sigma_2 \Delta L_j
    - L^{cT}_i i \sigma_2 \overline{\Delta} L_j^c \right)  \nonumber  \\
  & & +\, \frac 12 \left( \sigma_u H^T_u i\sigma_2 \overline{\Delta} H_u
    - \sigma_d H^T_d i\sigma_2 \Delta H_d \right) ,
\end{eqnarray}
where $H_u$ and $H_d$ are the light Higgs
doublets, $\sigma_{u,d} \equiv \sigma \alpha^2_{u,d}$ ($\alpha_{d,u}$
being defined by $H_{d,u}^{10}= \alpha_{d,u} H_{d,u}+\dots$) and
\begin{eqnarray}
  \Delta \equiv \frac{\vec \sigma}{\sqrt{2}} \cdot \vec{\Delta} =
    \left( \ba{cc}  \Delta^+ / \sqrt{2} & \Delta^{++}  \\
    \Delta^0 & - \Delta^+ / \sqrt{2}  \ea \right) ,
  & L_i = \left( \ba{c} \nu_i \\ e_i \ea \right) ,  \\
  \overline \Delta \equiv \frac{\vec \sigma}{\sqrt{2}} \cdot \vec{\overline{\Delta}} =
    \left( \begin{array}{cc}  \overline \Delta^- / \sqrt{2} & \overline \Delta^0  \\
    \overline \Delta^{--} & - \overline \Delta^- / \sqrt{2}  \end{array} \right) ,
  & L^c_i = \left( \ba{c} E^c_i \\ - N^c_i  \ea \right) .
\end{eqnarray}
The following superpotential terms will also be needed for the computation
of the CP asymmetry: 
\beq
W_{S,T}\ =\ f_{ij}\, L^{cT}_i\!
\left(\frac 12 \sqrt{\frac 35} S + \sqrt{\frac 12} T \right)\! i \sigma_2 L_j
+ \frac 12\, M_S S^2 +\frac 12\, M_T \mbox{Tr} (T^2) ~,
\label{eq:W_ST}
\eeq
where $S$ and $\vec{T}$ are the $(1,1,0)$ and $(1,3,0)$ components
of ${\bf 24}^{54}$, and $T \equiv \vec{\sigma} \cdot \vec{T} / \sqrt{2}$.

The chiral superfields  $\Delta$ and $\overline\Delta$ describe
two complex scalar fields, $\Delta_s$ and $\overline \Delta_s$,
and a Dirac spinor field $\Psi_\Delta$. 
Since we are dealing with a supersymmetric theory, we only need
to compute the CP asymmetry in the decays of one of the component
fields, e.g. the scalar triplet $\Delta_s$. This field has four decay modes
(neglecting phase space suppressed 3-body decays such as
$\Delta_s \rightarrow \tilde L^c\, \tilde e^c\, H_d$):
it can decay
into light leptons ($\Delta_s \rightarrow \overline{L}\, \overline{L}$), 
heavy sleptons ($\Delta_s \rightarrow \tilde{L}^c \tilde{L}^c$),
down-type Higgsinos
($\Delta_s \rightarrow \overline{\tilde{H}}_d \overline{\tilde{H}}_d$), 
and up-type Higgs bosons ($\Delta_s \rightarrow H_uH_u$).
The corresponding decay widths are given by
\beq
  \Gamma (\Delta_s \rightarrow a a)\ =\ \frac{M_\Delta}{32 \pi}\ \lambda^2_a\ ,
  \qquad \qquad a\ =\ \overline L,\ \tilde L^c,\ \overline{\tilde H}_d,\ H_u\ ,
\eeq
with
\beq
  \lambda^2_L\ \equiv\ \sum_{i,j=1}^3 |f_{ij}|^2\, ,  \quad
  \lambda^2_{L^c}\ \equiv\ \sum_{i,j=1}^{3} 
K \left(\frac{M^2_i}{M^2_\Delta},\frac{M^2_j}{M^2_\Delta}\right) |f_{ij}|^2\, ,  \quad
  \lambda^2_{H_{u,d}}\ \equiv\ |\sigma_{u,d}|^2\ .
\label{ls}
\eeq 
The kinematic factor $K$ for decays into $\tilde{L}^c$'s is
\beq
K(x_i,x_j) = \theta(1-\sqrt{x_i}-\sqrt{x_j})  \sqrt{\lambda(1,x_i,x_j)}\ ,
\eeq
where $\lambda(x,y,z)\equiv x^2+y^2+z^2-2 (xy + yz + xz)$.

The CP asymmetry in the decays of $\Delta_s$, $\Delta^*_s$ into light leptons
is defined by:
\beq
  \epsilon_{\Delta_s \rightarrow \overline{L}\,\overline{L}}\ 
    \equiv\ \frac{\Gamma({\Delta_s}\rightarrow \overline{L}\,\overline{L})
    - \Gamma(\Delta_s^*\rightarrow LL)}
    {\Gamma_{\Delta_s}+\Gamma_{\Delta_s^*}}~,
\label{epsiL}
\eeq
where $\Gamma_{\Delta_s} = M_\Delta \sum_a \lambda^2_a / (32 \pi)$
is the total $\Delta_s$ decay rate.
Unitarity and CPT imply $\Gamma_{\Delta_s} = \Gamma_{\Delta_s^*}$.
One can easily check that the asymmetries in the decays 
$\Delta_s\rightarrow \overline{\tilde{H}}_d\overline{\tilde{H}}_d$ 
and $\Delta_s\rightarrow H_uH_u$ vanish at the one loop order,
as they involve a single coupling $\sigma_d$ or $\sigma_u$. 
As a result, the equality $\Gamma_{\Delta_s} = \Gamma_{\Delta_s^*}$
reduces to
$\Gamma(\Delta_s^*\rightarrow LL)+\Gamma(\Delta_s^*\rightarrow 
\tilde{L}^{c*}\, \tilde{L}^{c*})=\Gamma(\Delta_s\rightarrow \overline{L}\,\overline{L})+
\Gamma(\Delta_s \rightarrow \tilde{L}^c\tilde{L}^c)$, i.e. the CP asymmetries
in $\Delta_s$ decays into light leptons and into heavy sleptons are exactly
opposite. This allows one to define
\beq
  \epsilon_{\Delta_s}\ \equiv\
    2\, \epsilon_{\Delta_s \rightarrow \tilde{L}^c\tilde{L}^c}\
    =\ - 2\, \epsilon_{\Delta_s \rightarrow \overline{L}\,\overline{L}}\ ,
\eeq
where the factor of $2$ accounts for the fact that two antileptons
are produced in each $\Delta_s$ decay.
Furthermore, supersymmetry ensures that the CP asymmetries
of all components of the $\Delta$, $\overline \Delta$ supermultiplets
are the same:
\beq
  \epsilon_{\Delta_s}\ =\ \epsilon_{\overline{\Delta}_s}\
    =\ \epsilon_{\Psi_\Delta}\ \equiv\ \epsilon_\Delta\ .
\eeq
%

\begin{figure}
\begin{center}
\begin{picture}(140,80)(-50,10)
\DashArrowLine(-100,50)(-70,50){5}
\ArrowLine(-40,80)(-70,50)
\ArrowLine(-40,20)(-70,50)
\Text(-95,42)[]{$\Delta_s$}
\Text(-55,22)[]{$L_i$}
\Text(-57,80)[]{$L_j$}
\Text(-57,51)[]{$f_{ij}$}
\DashArrowLine(-10,50)(30,50){5}
\ArrowLine(90,80)(60,80)
\ArrowLine(90,20)(60,20)
\Line(60,80)(60,20)
\DashArrowLine(30,50)(60,80){5}
\DashArrowLine(30,50)(60,20){5}
\Text(0,42)[]{$\Delta_s$}
\Text(38,30)[]{$\tilde{L}^c_k$}
\Text(38,75)[]{$\tilde{L}^c_l$}
\Text(73,50)[]{$S,T$}
\Text(102,80)[]{$L_j$}
\Text(102,20)[]{$L_i$}
\Text(44,51)[]{$f^*_{kl}$}
\Text(60,90)[]{$f_{lj}$}
\Text(60,10)[]{$f_{ki}$}
\end{picture}
\end{center}
\caption{Feynman diagrams responsible (together with the CP-conjugated
diagrams) for the CP asymmetry in the decays of the scalar triplet
into Standard Model lepton doublets.}
\label{deltas}
\end{figure}
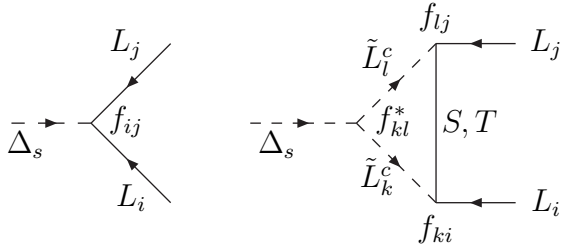

The Feynman diagrams relevant to the computation of
$\epsilon_\Delta$ are shown in Fig.~\ref{deltas}.
For arbitrary masses $M_\Delta$, $M_S$, $M_T$ and $M_i$
($i=1,2,3$), one obtains:
\beq
\epsilon_\Delta\
=\  \frac{1}{16\pi} \sum_{R=S,T} c_R 
\sum_{k,l=1}^{3} F\left(\frac{M^2_R}{M^2_\Delta},\frac{M^2_k}{M^2_\Delta},
\frac{M^2_l}{M^2_\Delta}\right)
\dfrac{{\rm Im}[f_{kl}(f^*ff^*)_{kl}]}
{32\pi\Gamma_{\Delta_s}/M_\Delta} ~,
\label{asy}
\eeq
where $c_S=3/5$ and $c_T=1$ are $SU(5)$ Clebsch-Gordan coefficients,
and the loop function $F$ reads:
\beq
F(x,x_k,x_l)\ =\ \theta(1 - \sqrt{x_k} - \sqrt{x_l})\,\sqrt{x}\,
\ln \left[\frac{1+2x-x_k-x_l+\sqrt{\lambda(1,x_k,x_l)}}
{1+2x-x_k-x_l-\sqrt{\lambda(1,x_k,x_l)}}\right] .
\eeq
It is instructive to consider some particular cases. In the case
$2 M_3 < M_\Delta$, all terms in the sum over $k, l$ in Eq.~(\ref{asy})
contribute to the asymmetry, and they add up to zero in the limit
of massless $\tilde L^c_i$'s ($M_i / M_{\Delta} \rightarrow 0$).
This is simply due to the fact that $\epsilon_\Delta \propto
\mbox{Im} \left[ \mbox{Tr} (f f^* f f^*) \right]$ in this limit, as discussed
previously. In the case $2 M_1 < M_\Delta < M_1 + M_2$, only
the term $k=l=1$ contributes and $F$ is maximal for $M_1=0$,
while it is reduced by about a factor of $2$ ($4$) for 
$2M_1/M_\Delta= 0.87~(0.97)$ and $M_{S,T} / M_\Delta \gtrsim 3$.
If in addition $M_S = M_T \equiv M_{24}$,
Eq.~(\ref{asy}) simplifies to, for $M_1 \ll M_\Delta$:
\beq
\epsilon_\Delta\
  \simeq\  \frac{1}{10\pi}\, \frac{M_{24}}{M_\Delta}\,
    \ln \left( 1 + \frac{M^2_\Delta}{M^2_{24}} \right) \frac{{\rm Im} [f_{11} (f^*ff^*)_{11}]}
    {\lambda^2_L + \lambda^2_{L^c_1} + \lambda^2_{H_u} + \lambda^2_{H_d}}~,
\label{asy_M1}
\eeq
where $\lambda_{L^c_1} \equiv |f_{11}|$.

\subsection{Other CP asymmetries \label{other}}

The CP asymmetries in the decays of the other components of the
(${\bf 15}$, $\bf \overline{15}$) multiplets can be computed
along the same lines, taking into account the following differences
with the triplet case (we consider the scalar components of the
$\Sigma$ and $Z$ multiplets for definiteness):
\begin{itemize}
  \item $\Sigma_s$ and $Z_s$ cannot decay into Higgs fields, since
the colour triplets contained in the ${\bf 10}$ Higgs multiplets
(to which the $\Sigma$ and $Z$ fields couple via the $SO(10)$ operator
$\bf 10\, 10\, 54$) must have GUT-scale masses in order to suppress
the dangerous $D=5$ proton decay operators. Thus, they have only two 
2-body decay modes, $\Sigma_s \rightarrow \overline{d^c}\, \overline{d^c}$
and $\Sigma_s \rightarrow \tilde D \tilde D$
(respectively $Z_s \rightarrow \overline L\, \overline{d^c}$ and
$Z_s \rightarrow \tilde L^c \tilde D$),
with opposite CP asymmetries:
  \bea
    \epsilon_\Sigma & \equiv &
      2\, \epsilon_{\Sigma_s \rightarrow \tilde D \tilde D}\
      =\ - 2\, \epsilon_{{\Sigma_s}\rightarrow \overline{d^c}\,\overline{d^c}}\ ,  \\
    \epsilon_Z & \equiv &
      \epsilon_{Z_s \rightarrow \tilde{L}^c \tilde D}\
      =\ -\, \epsilon_{{Z_s}\rightarrow \overline{L}\, \overline{d^c}}\ .
  \eea
  \item $\epsilon_\Sigma$ and $\epsilon_Z$
are given by a similar expression to Eq.~(\ref{asy}), with $M_\Delta$
replaced by $M_\Sigma$ and $M_Z$, respectively; however the
${\bf 24}^{54}$ components in the loop and the Clebsch-Gordan
coefficients are not the same. Furthermore, the total
decay rates $\Gamma_{\Sigma_s}$ and $\Gamma_{Z_s}$ differ from
$\Gamma_{\Delta_s}$ due to the absence of Higgs decay modes.
For $|\sigma| \gg |f_{ij}|$ this results in a significant enhancement
of $\epsilon_\Sigma$ and $\epsilon_Z$ with
respect to $\epsilon_\Delta$.
\end{itemize}

It is important to notice that, among the components of the
(${\bf 15}$, ${\bf \overline{15}}$) multiplets, only the $SU(2)_L$
triplets ($\Delta$, $\overline \Delta$) have ($B-L$)-violating interactions
(in the limit in which $M_{GUT}$-suppressed interactions are
neglected)\footnote{Indeed, once the colour triplets contained
in the ${\bf 10}$ Higgs multiplets have been integrated out,
the only renormalizable superpotential
terms involving $\Sigma$, $\overline \Sigma$, $Z$ or $\overline Z$ are
$\overline \Sigma \Sigma$, $\overline Z Z$, $\Sigma d^c d^c$,
$\overline \Sigma D D$, $Z L d^c$ and $\overline Z L^c D$.
These terms preserve $B-L$, as can be checked by assigning
suitable $B-L$ charges to the $\Sigma$ and $Z$ fields.
By contrast, the simultaneous presence
of the triplet interactions $\Delta L L$, $\overline \Delta H^{10}_u H^{10}_u$
and of the mass term $\Delta \overline \Delta$ violates $B-L$.}.
Hence the decays of the ($\Sigma$, $\overline \Sigma$)
and ($Z$, $\overline Z$) fields
cannot generate a sizeable $B-L$ asymmetry by themselves.
Still they produce asymmetries in the number densities of the species
$L_i$, $d^c_i$, $L^c_i$ and $D_i$,
which can affect the dynamical
evolution of the $B-L$ asymmetry generated in the decays of the
($\Delta$, $\overline \Delta$) components.

A similar statement can be made about the decays of the components of the
${\bf 24}^{54}$. Feynman diagrams
analogous to the ones shown in Fig.~\ref{deltas} generate a CP asymmetry
between, e.g., the decay $S_s \rightarrow \overline L\, \overline {L^c}$
and the CP-conjugated decay. However,
since all renormalizable interactions of the ${\bf 24}^{54}$ components
respect $B-L$,  their decays cannot generate 
a $B-L$ asymmetry by themselves.
Still they affect leptogenesis by producing asymmetries in the
number densities of the species $L_i$, $L^c_i$, $d^c_i$ and $D_i$.
In the following, we assume $M_{24} \gg M_\Delta$, so that either
the components of the ${\bf 24}^{54}$ are not efficiently produced in the thermal
bath after reheating, or the asymmetries generated in their decays
have been washed out before the components of the
($\Delta$, $\overline \Delta$) supermultiplets decay.

Finally, the right-handed neutrinos are a source of $B-L$
violation. In our scenario, they do not have standard Dirac couplings,
but they couple to the heavy leptons through the superpotential terms
$y_u \nu^c L^{16} H_u^{10}$ and $y_d \nu^c L^c H_d^{16}$.
If kinematically allowed, the associated two-body decays
will generate a $B-L$ asymmetry stored in the heavy (s)leptons,
before these decay to light species. In the following, we
assume that the right-handed neutrinos are heavier than the
($\Delta$, $\overline \Delta$) fields,
so that we can neglect their contribution to the final baryon
asymmetry -- either because they are not efficiently produced
after reheating, or because the asymmetries generated
in their decays have been washed out before the triplets decay.
This assumption is analogous to the one usually done
in type I leptogenesis, where the contribution of the next-to-lightest
right-handed neutrino to the lepton asymmetry is neglected.
Also, the decays of the heavy lepton fields into right-handed neutrinos,
which would introduce an additional source of $B-L$ violation
in our scenario, are kinematically forbidden.

\subsection{Dependence of the CP asymmetry on the light neutrino
mass parameters \label{subsec:dependence}}

The CP asymmetry in triplet decays, Eq.~(\ref{asy}), depends
on the heavy lepton masses and couplings, which in turn are related
to the low-energy lepton parameters.
Indeed, the superpotential terms $h_{ij} {\bf 16}_i {\bf 10}_j {\bf 16}$
yield the following GUT-scale mass relations:
\beq
  M_i\ =\ m_{e_i}\, \frac{v^{16}_1}{v^{16}_d}  \quad
    \left( =\, m_{d_i}\, \frac{v^{16}_1}{v^{16}_d} \right) ,
\label{eq:Mi}
\eeq
where $v_1^{16}$ is the vev of the $SU(5)$-singlet component
of $\bf 16$. Furthermore, it is possible to choose a ${\bf 10}_i$ basis
in which the charged lepton and heavy matter mass matrices
are simultaneously diagonal, and in this basis:
\beq
  f\ =\ \frac{2 M_\Delta}{\lambda_{H_u} v^2 \sin^2\! \beta}\
    U^* \mbox{Diag}\, (m_1, m_2, m_3) U^\dagger ,
\label{eq:f}
\eeq
where the $m_i$ are the light neutrino masses and $U$ is
the PMNS mixing matrix.
The non-renormalizable operators needed to correct 
the GUT-scale relations $m_\mu = m_s$ and $m_e = m_d$
will in general modify Eq.~(\ref{eq:Mi}),
but one can neglect this effect for an order-of-magnitude estimate
of the heavy lepton masses. With the $m_{e_i}$ evaluated
at the GUT scale, and assuming that the light Higgs doublet $H_d$
contains a large $H^{16}_d$ component (i.e. $v^{16}_d \sim v_d$),
one obtains:
\beq
  (M_1,\, M_2,\, M_3)\ \sim\ (2 \times 10^{11},\, 4 \times 10^{13},\,
    7 \times 10^{14})\ \mbox{GeV}\ \left( \frac{\tan \beta}{10} \right)
    \left( \frac{v^{16}_1}{10^{16}\, {\rm GeV}} \right) .
\eeq
Gauge coupling unification favours values of $v_1^{16}$ close
to $M_{GUT}$, unless the rank of $SO(10)$ is broken at the GUT scale
by the vev of a different Higgs multiplet (e.g. an extra ${\bf 16}$),
in which case $v_1^{16}$, hence the $M_i$, can be significantly smaller.

In the following,
we assume that the situation $M_1 \ll M_{\Delta} < M_1 + M_2$ is
realized\footnote{If instead $M_1 + M_2 < M_\Delta$,
$\epsilon_\Delta$ receives extra contributions, 
opening additional possibilities to achieve successful 
leptogenesis that could be analyzed along the same lines.}.
The CP asymmetry $\epsilon_\Delta$ is then given by
Eq.~(\ref{asy_M1}), which can be rewritten as:
\beq
\epsilon_\Delta\
  \simeq\  \frac{1}{10\pi}\ g\! \left( \frac{M^2_{24}}{M^2_\Delta} \right) \lambda^2_L\
  \frac{\lambda^2_L}{\lambda^2_L + \lambda^2_{L^c_1} + \lambda^2_{H_u} + \lambda^2_{H_d}}\
   \frac{{\rm Im} [m_{11} (m^*mm^*)_{11}]}{\overline m^4}\ ,
\label{asy_m}
\eeq
where $g (x) \equiv \sqrt{x}\, \ln\, [1 + 1/ x]$,
$m \equiv U^* \mbox{Diag}\, (m_1, m_2, m_3) U^\dagger$ and
$\overline m^2 \equiv \sum_i m^2_i$.
In writing Eq.~(\ref{asy_m}), we assumed the absence of a mismatch
between the mass eigenstate bases of charged leptons and of
the heavy lepton fields, as in Eq.~(\ref{eq:f}). Such a mismatch may
arise from the corrections needed to account for $m_{d_i} \neq m_{e_i}$,
but Eq.~(\ref{asy_m}) is still a good approximation
if the unitary matrix that describes this mismatch is characterized by small
mixing angles (compared with the uncertainties on the PMNS angles).

The factor in Eq.~(\ref{asy_m}) that explicitly depends on the light neutrino
mass parameters reads:
\bea
  \frac{{\rm Im} [m_{11} (m^*mm^*)_{11}]}{\overline m^4}\
     =\, -\, \frac{1}{\overline m^4}\, \biggl\{ c^4_{13} c^2_{12} s^2_{12}
    \sin (2 \rho)\, m_1 m_2 \Delta m^2_{21}  \hskip 3cm  \nonumber  \\ 
  + c^2_{13} s^2_{13} c^2_{12} \sin 2 (\rho -\sigma)\, m_1 m_3 \Delta m^2_{31}
    - c^2_{13} s^2_{13} s^2_{12} \sin (2 \sigma)\, m_2 m_3 \Delta m^2_{32} \biggl\}\ ,
\label{eq:Im}
\eea
where $c_{ij} \equiv \cos \theta_{ij}$, $s_{ij} \equiv \sin \theta_{ij}$,
$\Delta m^2_{ji} \equiv m^2_j - m^2_i$, and we adopted the parametrization
$U_{ei} = (c_{13} c_{12} e^{i\rho}$, $c_{13} s_{12}$, $s_{13} e^{i\sigma})$,
in which $\rho$ and $\sigma$ are the two Majorana-type CP-violating phases
to which neutrinoless double beta decay is sensitive. Indeed,
neutrinoless double beta decay depends on the effective Majorana mass
$|m_{ee}| = |\sum_i m_i U^2_{ei}|$.

Some comments about Eq.~(\ref{asy_m}) are in order.
First, to the extent discussed above, the CP asymmetry
is predicted in terms of the light neutrino parameters
once $\tan \beta$ and the flavour-blind parameters $\lambda_L$,
$\lambda_{H_u}$, $\lambda_{H_d}$ and $M_\Delta / M_{24}$
are specified ($\lambda_{L^c_1}$ is not an independent
parameter, since it is related to $\lambda_L$ by 
$\lambda_{L^c_1} = \lambda_L |m_{11}| / \overline m$).
This is a noticeable difference with leptogenesis in the standard
type I and type II seesaw mechanisms. Second, the CP asymmetry
depends on the same two CP-violating phases as neutrinoless
double beta decay; observing CP violation in neutrino oscillations
would not be enough to test the validity of the present scenario.
Finally, it is very sensitive to the yet unknown value of $\theta_{13}$
and to the type of mass spectrum.

\begin{figure}
\begin{center}
\includegraphics[width=0.95\textwidth]{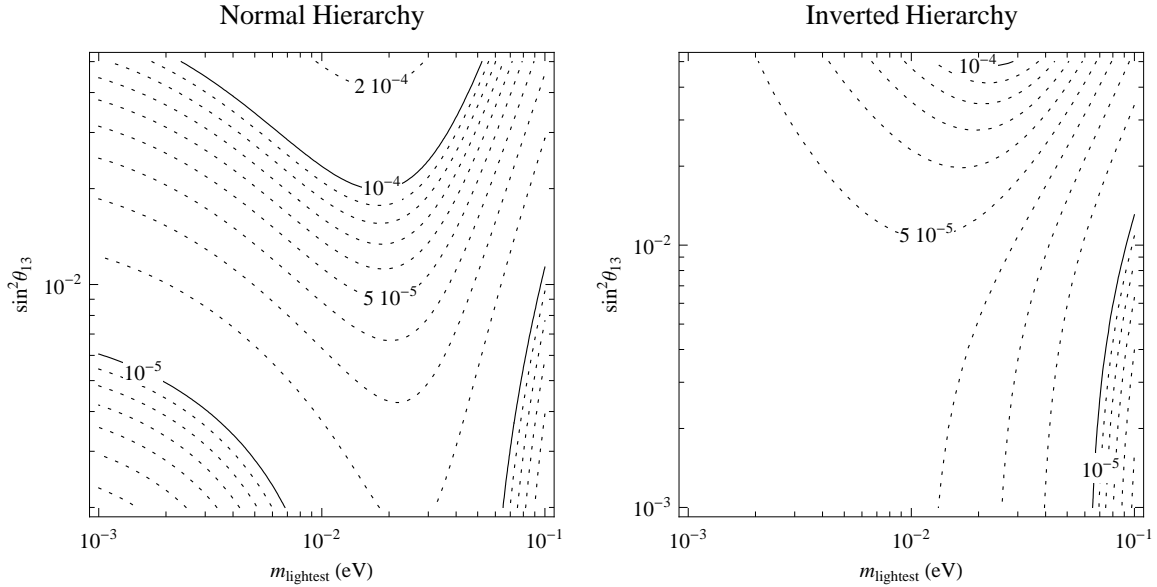}
\end{center}
\caption{The CP-asymmetry $\epsilon_\Delta$ in units of $\lambda^2_L$
as function of the lightest neutrino mass and of $\sin^2\theta_{13}$
in the cases of normal and inverted hierarchy of light neutrino masses.
The parameters $|\Delta m^2_{32}|$,
$\Delta m^2_{21}$ and $\theta_{12}$ are chosen as specified in the text.
The asymmetry is maximized with respect to the CP-violating phases
and to the ratio $M_\Delta/M_{24}$.}
\label{fig:maximal}
\end{figure}

To illustrate this point, let us look for the parameter values that
maximize the quantity in Eq.~(\ref{eq:Im}). If $\theta_{13}$ is close
to its present experimental upper bound, $\sin^2 \theta_{13} \approx 0.05$,
the maximal value of Eq.~(\ref{eq:Im}) is obtained for a normal
neutrino mass hierarchy with $m_1 \approx m_2 \approx 0.01$ eV
and $\rho = 0$, $\sigma = \pi / 4$, and it is given by
$\approx c^2_{13} s^2_{13} \sqrt{\Delta m^2_{21} / \Delta m^2_{32}}$.
A more precise estimate gives $9.2 \times 10^{-3}$
for $m_1 = 0.016$ eV and $s^2_{13} = 0.05$.
For $\sin^2 \theta_{13} \lesssim 0.01$, the maximum value
of Eq.~(\ref{eq:Im}) is obtained for an inverted hierarchy
with $m_3 = 0$ and $\rho = - \pi / 4$, and it is given by
$\approx c^2_{12} s^2_{12} \Delta m^2_{21} / (4 |\Delta m^2_{32}|)$.
A more precise estimate gives $1.7 \times 10^{-3}$ for $\theta_{13} = 0$.
Taking $\lambda^2_L \gg \lambda^2_{H_u}, \lambda^2_{H_d}$
and choosing $M_{24} \simeq M_\Delta / 2$ in order to maximize the loop
function $g (M^2_{24} / M^2_\Delta)$, one obtains, for the above two sets
of light neutrino mass parameters:
\bea
  \epsilon_\Delta & \simeq & 2.2 \times 10^{-4}\, \lambda^2_L \qquad \quad
    ({\rm maximum}\, \theta_{13})\ ,  \label{epsilon_max}  \\
  & \simeq &  3.4 \times 10^{-5}\, \lambda^2_L \qquad \quad
    ({\rm vanishing}\, \theta_{13})\ ,
\eea
where the value of $\lambda^2_L$ is bounded by perturbativity
(requiring $|f_{ij}| \leq 1$, one can take $\lambda^2_L$ as large
as $5$ for a normal hierarchical spectrum).
In the above estimates, we used
$|\Delta m^2_{31}| = 2.4 \times 10^{-3}\, {\rm eV}^2$,
$\Delta m^2_{21} = 7.6 \times 10^{-5}\, {\rm eV}^2$
and $\sin^2 \theta_{12} = 0.32$~\cite{Schwetz07}.
The asymmetry, maximized with respect to the CP-violating phases
and to $M_\Delta/M_{24}$, is plotted in \Fig{maximal} as a function
of the lightest neutrino mass and $\theta_{13}$, for the cases
of normal and inverted hierarchy.

Eq.~(\ref{epsilon_max}), together with Eq.~(\ref{bsr}), shows that
an efficiency factor as small as $10^{-5} - 10^{-4}$ is sufficient
for successful
baryogenesis if $\epsilon_\Delta$ is close to its maximum value.
A quantitative estimate of $\eta$ requires in general the numerical
resolution of the Boltzmann equations, and is beyond the scope
of the present paper\footnote{A numerical analysis of the efficiency
has been performed in Refs.~\cite{hrs,stoc} for the simpler case
of the Standard Model augmented with an $SU(2)_L$ Higgs triplet,
and in Ref.~\cite{CS06} for its supersymmetric extension.}.
However, the region of the parameter space which leads to a large
efficiency factor can be discussed analytically, as we do
in Section~\ref{sec:efficiency}.

\section{Dynamical evolution of the $B-L$ asymmetry\label{sec:efficiency}}

As discussed in Section~\ref{other}, in the limit in which one
neglects $M_{GUT}$-suppressed interactions, the only source
for the $B-L$ asymmetry in our scenario
are the out-of-equilibrium decays of the components
of the ($\Delta$, $\overline \Delta$) supermultiplets.
In principle, the $M_{GUT}$-suppressed interactions could be relevant
for the subsequent decays of $L^c_1$ and $D_1$, which would then
violate $B-L$ and affect the final baryon asymmetry.
However, it turns out that the dominant decay modes of $L^c_1$
and $D_1$ preserve $B-L$ as long as the light Higgs doublet $H_d$
contains a non-negligible $H^d_{10}$ component\footnote{Unless
$\alpha_d$ is very small, the (B-L)-conserving decay modes,
even though suppressed by small Yukawa couplings, dominate over the
(B-L)-violating ones mediated by the heavy right-handed neutrinos
or by the colour triplets. For instance, the dominant decay mode
of $\tilde L^c_1$ (respectively $\tilde D_1$) is
$\tilde L^c_1 \rightarrow \tilde e^c H_d$ (respectively
$\tilde D_1 \rightarrow \tilde Q H_d$)
over most of the parameter space.
It is interesting to note that natural realizations of the doublet-triplet
splitting seem to require $\alpha_d \neq 0$ in our scenario. \label{foo:Lc_decays}},
and we shall assume that this is the case in the following.

As the $B-L$ asymmetry originates from the triplet decays,
we expect that a good estimate of the efficiency can be
obtained by considering the decays and interactions of the
($\Delta$, $\overline \Delta$) fields only, as we do below
in order to simplify the discussion.
We shall check in
Section~\ref{subsec:check} that including all components
of the ($\bf 15$, $\bf \overline {15}$) multiplets in the analysis
does not affect our conclusions.
We therefore define the efficiency factor $\eta$ as follows:
\beq
\frac{n_{B-L}}{s}\
=\ \eta\, \epsilon_\Delta 
\left[\frac{n^{eq}_{\Delta_s+\Delta_s^*}}{s}+
\frac{n^{eq}_{\overline{\Delta}_s+\overline{\Delta}_s^*}}{s}+
\frac{n^{eq}_{\Psi_\Delta+\overline{\Psi}_\Delta}}{s}\right]_{T\gg M_\Delta} ~,
\eeq
where $s = (2 \pi^2 / 45) g_{\star S} T^3$ is the entropy density.
After conversion of the $B-L$ asymmetry by the $(B+L)$-violating
sphalerons, the final baryon asymmetry reads:
\beq
\frac{n_B}{s}\ =\ \frac{8}{23} \frac{n_{B-L}}{s}\ =\ 
7.62 \times 10^{-3}\, \eta\,  \epsilon_\Delta ~,
\label{bsr}
\eeq
where we have used $g_{\star S} = g_{\star S} (MSSM)
+ g_{\star S} ({\bf 5}_1, {\bf \overline 5}_1) = 266.25$
(we assume as in the previous section that a single (${\bf 5}^{10}_M$,
${\bf \overline 5}^{16}_M$) pair is lighter than the triplet,
namely $M_1 \ll M_\Delta < M_1 + M_2$).
The observed value of the baryon-to-entropy ratio,
$(n_B/s)_{WMAP}= (8.82\pm0.23) \times 10^{-11}$ \cite{WMAP},
requires $\eta\, \epsilon_{\Delta} \simeq 10^{-8}$.

\subsection{Boltzmann equations}

Since the interaction rates and CP asymmetries involving different
components of the ($\Delta$, $\overline \Delta$) superfields are
related by supersymmetry, one can obtain a valid estimate of $\eta$
by considering solely the scalar triplet $\Delta_s$. The problem of
computing $\eta$ is then similar to the case of the Higgs triplet extension
of the Standard Model studied in Ref.~\cite{hrs}, with however two
important differences: our scenario involves another species carrying
lepton number to which $\Delta_s$ can decay ($\tilde L^c_1$),
and there is no CP asymmetry in the Higgs/Higgsino decay channels.
The relevant Boltzmann equations
(written for simplicity in the $\alpha_d = 0$ case, in which $\Delta_s$
has no Higgsino decay mode) read:
\begin{eqnarray}
  s H z \frac{d \Sigma_{\Delta_s}}{dz} & = & - \gamma_D
    \left( \frac{\Sigma_{\Delta_s}}{\Sigma^{eq}_{\Delta_s}} - 1 \right) - 2 \gamma_A
     \left( \left(\frac{\Sigma_{\Delta_s}}{\Sigma^{eq}_{\Delta_s}}\right)^2 - 1 \right) , 
     \label{eq:BE_SigmaT}  \\
  s H z \frac{d \Delta_{\Delta_s}}{dz} & = & - \gamma_D \left( \frac{\Delta_{\Delta_s}}{\Sigma^{eq}_{\Delta_s}}
    + B_L \frac{\Delta_L}{Y^{eq}_L} - B_{H_u} \frac{\Delta_{H_u}}{Y^{eq}_{H_u}}
    - B_{L^c_1} \frac{\Delta_{\tilde L^c_1}}{Y^{eq}_{\tilde L^c_1}} \right) ,
     \label{eq:BE_T}  \\
  s H z \frac{d \Delta_L}{dz} & = & \gamma_D \epsilon_\Delta
    \left( \frac{\Sigma_{\Delta_s}}{\Sigma^{eq}_{\Delta_s}} - 1 \right) - 2 \gamma_D B_L
    \left( \frac{\Delta_L}{Y^{eq}_L} + \frac{\Delta_{\Delta_s}}{\Sigma^{eq}_{\Delta_s}} \right) ,
     \label{eq:BE_L}  \\
  s H z \frac{d \Delta_{H_u}}{dz} & = & - 2 \gamma_D B_{H_u}
    \left( \frac{\Delta_{H_u}}{Y^{eq}_{H_u}} - \frac{\Delta_{\Delta_s}}{\Sigma^{eq}_{\Delta_s}} \right) ,
     \label{eq:BE_Hu}  \\
  s H z \frac{d \Delta_{\tilde{L}^c_1}}{dz} & = & \gamma_D \epsilon_\Delta
    \left( \frac{\Sigma_{\Delta_s}}{\Sigma^{eq}_{\Delta_s}} - 1 \right) - 2 \gamma_D B_{L^c_1}
    \left( \frac{\Delta_{\tilde L^c_1}}{Y^{eq}_{\tilde L^c_1}}
    - \frac{\Delta_{\Delta_s}}{\Sigma^{eq}_{\Delta_s}} \right) ,  \label{eq:BE_Lc}
\end{eqnarray}
where $z = M_\Delta / T$, $Y_X = n_X / s$,
$\Delta_X \equiv Y_X - Y_{\overline X}$ is the asymmetry stored
in the species $X$, and $\Sigma_{\Delta_s} \equiv Y_{\Delta_s} + Y_{\Delta^*_s}$
is the total density of $\Delta_s$ and $\Delta^*_s$;
$\gamma_D = s \Sigma^{eq}_{\Delta_s} \Gamma_{\Delta_s} K_1(z )/ K_2(z)$
(where $K_1(z)$ and $K_2(z)$ are modified Bessel functions)
and $\gamma_A$ are the space-time densities of the ($\Delta_s$, $\Delta^*_s$)
decays and of the gauge scatterings $\Delta_s \Delta^*_s \rightarrow$ 
{\it lighter particles}, respectively; $B_L$, $B_{H_u}$ and $B_{L^c_1}$
are the branching ratios of $\Delta_s$ decays into $\overline L\, \overline L$,
$H_u H_u$ and $\tilde L^c_1 \tilde L^c_1$, respectively.
Note that we included only decays, inverse decays and gauge scatterings
in Eqs.~(\ref{eq:BE_SigmaT}) to (\ref{eq:BE_Lc}). In particular, we omitted
the triplet-mediated $\Delta L = 2$ scatterings
$L L \leftrightarrow H^*_u H^*_u$
and $L H_u \leftrightarrow \overline L H^*_u$,
which due to the smallness of neutrino masses are much slower that the
expansion of the Universe  (except for very large values of $M_\Delta$).
Scatterings such as $\tilde L^c_1 \tilde L^c_1 \leftrightarrow H_u H_u$
($\Delta L = 2$, triplet-mediated) and $\tilde L^c_1 \tilde L^c_1 \leftrightarrow 
\overline L\, \overline L$ ($\Delta L = 0$, triplet- or ($S$, $T$)-mediated)
are even slower in the region of the parameter space that we shall
consider below.
We also neglected the washout due to the inverse decays 
of $S$ and $T$, which are Boltzmann suppressed at $T \sim M_{\Delta}$
since $M_{\Delta} \ll M_{24}$.

An important point to notice is that the Boltzmann equations~(\ref{eq:BE_T})
to~(\ref{eq:BE_Lc}) are not linearly independent. As a result, the following
combination of asymmetries is preserved:
\beq
  2 \Delta_{\Delta_s} - \Delta_L + \Delta_{H_u} + \Delta_{\tilde L^c_1}\ =\ 0\ .
\label{eq:sum_rule}
\eeq
This relation generalizes the sum rule of Ref.~\cite{hrs}.

\subsection{Conditions for an order one efficiency}

In triplet leptogenesis, an order one efficiency can be obtained
even though the total triplet decay rate is larger than the expansion rate
of the universe, provided that one of the decay channels is out of equilibrium
and sufficiently decoupled from the fast channel(s). This has been first
pointed out in Ref.~\cite{hrs}, where the case of the Standard Model augmented
with a scalar triplet has been studied. A similar conclusion can be reached
in our scenario, with the role of the slow decay mode played by
$\Delta_s \rightarrow \tilde L^c_1 \tilde L^c_1$,
as we now discuss.

Let us first consider the out-of-equilibrium conditions
for the various decay channels.
The condition for the decay $\Delta_s \rightarrow a a$
($a = \overline L, H_u, \tilde L^c_1$) to be out of equilibrium 
at $T = M_\Delta$ is $K_a \equiv \Gamma (\Delta_s \rightarrow a a)
/ H(M_\Delta) \ll 1$. Since
$\Gamma (\Delta_s \rightarrow a a) = \lambda^2_a M_\Delta / (32 \pi)$
and $H (T) = 1.66 \sqrt{g_\star}\, T^2 / M_P$ (with $g_\star = g_{\star S}
= 266.25$ during leptogenesis), this condition translates into:
\beq
  \lambda_a\ \ll\ 10^{-2}\, \sqrt{\frac{M_\Delta}{10^{12}\, {\rm GeV}}}\ ,
\label{ooe}
\eeq
where we recall that $\lambda_L = \sqrt{{\rm Tr}\, (ff^*)}$, $\lambda_{H_u} = |\sigma_u|$,
and $\lambda_{L^c_1} = |f_{11}|$. Due to $\bar m^2 \equiv \sum_i m^2_i
\propto \lambda^2_L \lambda^2_{H_u} / M^2_\Delta$,
the product $K_L K_{H_u}$ is controlled by the scale of neutrino masses:
\beq
  K_LK_{H_u}\ \simeq\ \frac{220}{\sin^4\! \beta}\,
    \left( \frac{\overline m}{0.05\, \mbox{eV}} \right)^2 .
\label{mali}
\eeq
Hence, at least one of the two channels $\Delta_s \rightarrow \overline L\, \overline L$
and $\Delta_s \rightarrow H_u H_u$ (easily both) must be in equilibrium. Moreover,
we have $\lambda_{L^c_1} < \lambda_L$ (with
$\lambda_{L^c_1} \lesssim 0.2\, \lambda_L$ for hierarchical light
neutrino masses). The candidate out-of-equilibrium decay is therefore
$\Delta_s \rightarrow \tilde L^c_1 \tilde L^c_1$.
As we are going to see, a large efficiency can be reached in the region
$\lambda_{L^c_1} \ll \lambda_{H_u}, \lambda_L$, in which triplet decays
into leptons and Higgs are in equilibrium\footnote{One may wonder
what would happen if either $K_{H_u} \ll 1$
or $K_L \ll 1$. If $K_{H_u} \ll 1$, only a small $\Delta_{H_u}$
is generated via Eq.~(\ref{eq:BE_Hu}), resulting in a suppressed $Y_{B-L}$
according to Eq.~(\ref{eq:YB_Hu}). If $K_L \ll 1$, one has in particular
$\lambda_L \ll \lambda_{H_u}$ and this implies a strong
suppression of $\epsilon_\Delta$ according to Eq.~(\ref{asy_m}).}
($K_{H_u}, K_L \gtrsim 1$), while
decays into $\tilde L^c_1 \tilde L^c_1$ are not ($K_{L^c_1} \ll 1$).

Gauge scatterings ($\Delta_s \Delta^*_s \rightarrow$ {\it lighter particles})
first create an equilibrium population of triplets and antitriplets.
During leptogenesis, this population is kept close to thermal equilibrium
by decays and inverse decays, thanks to Eq.~(\ref{mali}) (the fact
that $\gamma_A < \gamma_D$ for $T < M_\Delta$ allows the triplets to decay
before annihilating~\cite{hrs}).
In spite of this, a large asymmetry $\Delta_{\tilde L^c_1}$ can develop
without being washed out, due to the fact that $K_{L^c_1} \ll 1$.
This implies an asymmetry between the abundances of triplets and
antitriplets, which is then transferred to $\Delta_L$ and $\Delta_{H_u}$
through their decays. After all triplets have decayed,
and before the $\tilde L^c_1$'s decay,
we end up with:
\beq
  Y_{B-L}\ =\ \Delta_{\tilde L^c_1} - \Delta_L\ =\ - \Delta_{H_u}\ ,
\label{eq:YB_Hu}
\eeq
where we made use of the sum rule~(\ref{eq:sum_rule}).
In order to create a large $B-L$ asymmetry, a large $\Delta_{H_u}$
is thus needed. We can check that a large $\Delta_{H_u}$ indeed
forms, and relate its size to $\Delta_{\tilde L^c_1}$,
by noticing that the combinations
\beq
  \frac{\Delta_L}{Y^{eq}_L} + \frac{\Delta_{\Delta_s}}{\Sigma^{eq}_{\Delta_s}}
  \qquad \mbox{and} \qquad
  \frac{\Delta_{H_u}}{Y^{eq}_{H_u}} - \frac{\Delta_{\Delta_s}}{\Sigma^{eq}_{\Delta_s}}\ ,
\label{eq:combinations}
\eeq
which multiply ($- 2 \gamma_D B_L$) in Eq.~(\ref{eq:BE_L}) and
($- 2 \gamma_D B_{H_u}$) in Eq.~(\ref{eq:BE_Hu}), respectively,
are forced to vanish due to $K_L$, $K_{H_u} \gtrsim 1$. One can check
that this is still the case after most triplets have decayed. Then, using
again the sum rule~(\ref{eq:sum_rule}), we obtain:
\beq
  Y_{B-L}\
    \simeq\ \frac{Y^{eq}_{H_u}}{Y^{eq}_L + Y^{eq}_{H_u}}\
    \Delta_{\tilde L^c_1}\ =\ \frac{4}{7}\, \Delta_{\tilde L^c_1}\ .
\label{eq:Y_B-L}
\eeq
To estimate $\Delta_{\tilde L^c_1}$, let us note that, in the limit
$\gamma_A = B_{L^c_1} = 0$ (remember that
$\gamma_A \ll \gamma_D$ during leptogenesis), Eqs.~(\ref{eq:BE_SigmaT})
and~(\ref{eq:BE_Lc}) give $\Delta_{\tilde L^c_1}
= \epsilon_\Delta \Sigma^{eq}_{\Delta_s} (T\gg M_\Delta)$. 
Taking into account the effect of gauge scatterings and the washout
by inverse decays introduces an order one factor $\eta_0 < 1$ such that
\beq
  \Delta_{\tilde L^c_1}\ =\ \eta_0 \epsilon_\Delta
    \Sigma^{eq}_{\Delta_s} (T\gg M_\Delta)\ .
\label{eq:Delta_Lc}
\eeq
The precise value of $\eta_0$ must be determined by solving
numerically the complete set of Boltzmann equations, but the fact that
$\gamma_A \ll \gamma_D$ and $K_{L^c_1} \ll 1$
guarantees that it cannot be much smaller than 1.
We conclude that an order one efficiency $\eta \simeq 4 \eta_0 / 7$
can be reached in the region of parameter space where
$\lambda_{L^c_1} \ll \lambda_{H_u}, \lambda_L$.
This region will be identified more precisely in Section~\ref{subsec:discussion}.
Although we set $\alpha_d = 0$ in the Boltzmann equations for simplicity,
this value of $\eta / \eta_0$ is valid for non-vanishing values of $\alpha_d$
such that $K_{H_d} \ll 1$ (let us recall that a non-negligible
$\alpha_d$ guarantees that the dominant $\tilde L^c_1$
decay modes preserve $B-L$).
For values of $\alpha_d$ such that $K_{H_d} \gtrsim 1$, one obtains
$\eta \simeq 7 \eta_0 / 10$.

Let us add a comment on lepton flavour effects. In the above,
we worked in the ``one-flavour approximation'', i.e. we assumed
a single generation of (light) leptons. Including the lepton flavours
would greatly complicate the above discussion; however we can
estimate their potential effect by assuming
$\Delta_{L_e} \approx \Delta_{L_\mu} \approx \Delta_{L_\tau}$,
which is not unreasonable given the mild hierarchy between the
decay rates $\Delta_s \rightarrow \overline L_i \overline L_j$.
In this case, the ``flavoured'' Boltzmann equations reduce
to Eqs.~(\ref{eq:BE_SigmaT})~--~(\ref{eq:BE_Lc}) with
$Y^{eq}_L$ replaced by $3 Y^{eq}_L$. This leads to an efficiency
$\eta \simeq 4 \eta_0 /13$,
to be compared with $\eta \simeq 4 \eta_0 / 7$ in the one-flavour case
(keeping in mind that the value of $\eta_0$ itself depends on whether
flavour effects are taken into account or not).

\subsection{Effect of the other components of ($\bf 15$, $\bf \overline{15}$)
\label{subsec:check}}

Let us now discuss how the above results are modified when
all components of the ($\bf 15$, $\bf \overline{15}$) multiplets
are taken into account. In this case,
Eqs.~(\ref{eq:BE_SigmaT})~--~(\ref{eq:BE_Lc}) are replaced
with a larger number of Boltzmann
equations describing the evolution of $\Sigma_{\Delta_s}$,
$\Sigma_{\Sigma_s}$, $\Sigma_{Z_s}$ and of the asymmetries stored
in the species $\Delta_s$, $\Sigma_s$, $Z_s$, $L$, $d^c$, $H_u$,
$\tilde L^c_1$ and $\tilde D_1$. Instead of the sum rule
(\ref{eq:sum_rule}), we now find two relations between asymmetries:
\begin{eqnarray}
  2 \Delta_{\Sigma_s} + \Delta_{Z_s} + \Delta_{\tilde D_1}
    - \Delta_{d^c} & = & 0\ ,
\label{eq:sum_rule2}  \\
  2 \Delta_{\Delta_s} + \Delta_{Z_s} + \Delta_{\tilde L^c_1} - \Delta_L
    + \Delta_{H_u} & = & 0\ .
\label{eq:sum_rule3}
\end{eqnarray}
The first relation tells us that $\Delta_{\tilde D_1} = \Delta_{d^c}$
after all components of the ($\bf 15$, $\bf \overline{15}$) have decayed,
hence the coloured species do not contribute to the $B-L$
asymmetry. This is consistent with the fact that all relevant
interactions preserve the baryon number. The second relation is analogous
to Eq.~(\ref{eq:sum_rule}). Furthermore, the combinations of asymmetries
(\ref{eq:combinations}) are still driven to zero by the Boltzmann equations.
As a result Eq.~(\ref{eq:Y_B-L}) still holds, but now both
the ($\Delta_s$, $\Delta^*_s$) and the ($Z_s$, $Z^*_s$) decays
contribute to $\Delta_{\tilde L^c_1}$.
In the limit where $\gamma^\Delta_A = \gamma^Z_A
= B (\Delta_s \rightarrow \tilde L^c_1 \tilde L^c_1)
= B (Z_s \rightarrow \tilde L^c_1 \tilde D_1) = 0$,
one obtains 
 $\Delta_{\tilde L^c_1}
 = \epsilon_\Delta \Sigma^{eq}_{\Delta_s} (T\gg M_\Delta)
+ \epsilon_Z \Sigma^{eq}_{Z_s} (T\gg M_Z)
= (\epsilon_\Delta + 2 \epsilon_Z) \Sigma^{eq}_{\Delta_s} (T\gg M_\Delta)$.
Taking into account the effect of gauge scatterings and the washout
by inverse decays introduces an order one factor $\eta_0 < 1$ such that
\beq
  \Delta_{\tilde L^c_1}\ =\ \eta_0 (\epsilon_\Delta + 2 \epsilon_Z)
    \Sigma^{eq}_{\Delta_s} (T\gg M_\Delta)\ ,
\eeq
where, as discussed in Section~\ref{other}, $\epsilon_Z$ and
$\epsilon_\Delta$ differ by an order one coefficient but have
the same sign. This leads us to define the efficiency factor $\eta$ as
\beq
  Y_{B-L}\ =\ \eta (\epsilon_\Delta + 2 \epsilon_Z)
    \Sigma^{eq}_{\Delta_s} (T\gg M_\Delta)\ ,
\eeq
so that $\eta \simeq 4 \eta_0 / 7$ for $K_{H_d} \ll 1$ (respectively
$\eta \simeq 7 \eta_0 / 10$ for $K_{H_d} \gtrsim 1$). Comparing the above
equations with Eqs.~(\ref{eq:Y_B-L}) and~(\ref{eq:Delta_Lc}) shows
an enhancement of $Y_{B-L}$ by a factor 
$|(\epsilon_\Delta + 2 \epsilon_Z) / \epsilon_\Delta|$, which may be
compensated by a smaller $\eta_0$, since $\Delta_{\tilde L^c_1}$
is washed out by the inverse decays of both $\Delta_s$ and $Z_s$.
Therefore, we do not expect the final value of the baryon asymmetry
to be significantly affected by the presence of the
($\Sigma$, $\overline \Sigma$) and ($Z$, $\overline Z$) fields.

\subsection{Discussion \label{subsec:discussion}}

The conditions for an order one efficiency,
\beq
  K_{L^c_1} \ll 1\ , \quad K_L, K_{H_u} \gtrsim 1 \quad {\rm and} \quad
    M_{24} \gg M_\Delta\ ,
\eeq
have some impact on the value of the CP asymmetry, which is not allowed
to be maximal anymore. First, the $K_{L^c_1} \ll 1$ condition sets
an upper bound on the absolute value of the $m_{11}$ factor in
Eq.~(\ref{asy_m}), which also enters the neutrinoless double beta
decay rate. Such a bound can only be satisfied for a normal hierarchy
of the light neutrino mass spectrum. 
Furthermore, the condition $K_{H_u} \gtrsim 1$ prevents us from taking
a large value of $\lambda_L$, since Eq.~(\ref{mnu}) implies
$\lambda^2_L \simeq 0.05\, K^{-1}_{H_u} (M_\Delta / 10^{12}\, \mbox{GeV})
(\overline m / 0.05\, {\rm eV})^2$. Finally, avoiding the washout from
($S$, $T$)-related processes requires $M_\Delta \ll M_{24}$, which prevents
the loop function from taking its maximal value.
As a result, $\epsilon_{\Delta}$ takes rather small values in the region
of parameter space where the efficiency is of order one, as we discuss
below in greater detail. 

Let us first recall that, within the uncertainties discussed
in Section~\ref{subsec:dependence}, we can take the flavour-blind quantities
$\lambda_L$, $\lambda_{H_u}$, $\lambda_{H_d}$ and $M_\Delta/M_{24}$
as independent parameters (in addition to $\tan\beta$, the light neutrino
masses and the PMNS matrix, which in principle can all  have an independent
experimental determination).
For definiteness, we consider as before the small $\alpha_d$ case,
so that we can in practice set $\lambda_{H_d} = 0$, and we choose
$\tan \beta = 10$ (these parameters do not play a crucial role).
We also set $M_\Delta/M_{24} = 0.1$, and observe
that $\epsilon_{\Delta}$ scales linearly with $M_\Delta/M_{24}$
for the necessarily small values of this ratio. We are then left with
$\lambda_L$, $\lambda_H\equiv\lambda_{H_u}$ and the light neutrino parameters.

\begin{figure}[t]
\begin{center}
\includegraphics[width=0.55\textwidth]{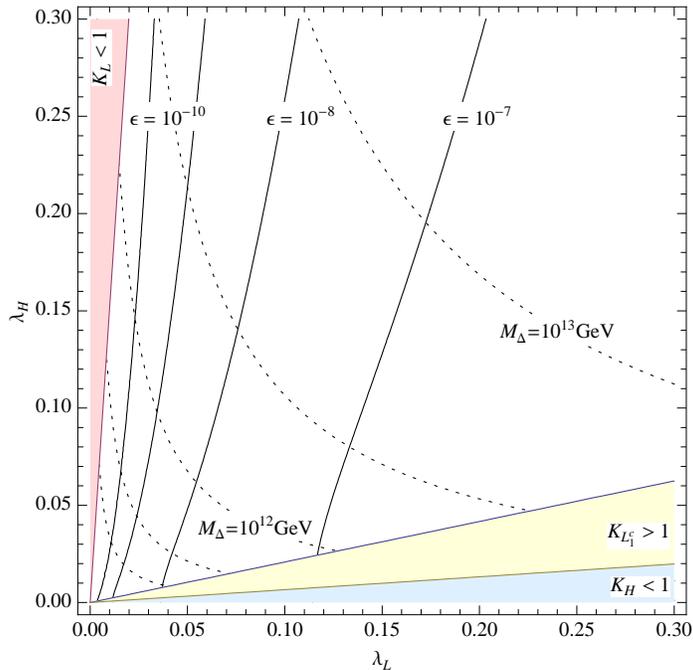}
\end{center}
\caption{Contours of constant $\epsilon_\Delta$ and $M_\Delta$
in the $\lambda_L$--$\lambda_H$ plane for a normal hierarchical neutrino mass
spectrum with $m_1\ll m_2$, $\sin^2\theta_{13} = 0.05$ and $\sin 2\sigma = 1$.
In the shaded regions, the conditions for a large efficiency are not satisfied.}
\label{fig:parspace}
\end{figure}

Contour lines of constant $\epsilon_\Delta$ in the
$\lambda_L$--$\lambda_H$ plane are shown in \Fig{parspace}
for the case of a normal hierarchical spectrum with $m_1\ll m_2$.
In this limit, the asymmetry scales as $\sin^2\theta_{13}$
and $\sin 2\sigma$, and is therefore maximized by
$\sin^2\theta_{13} = \sin^2\theta_{13}^\text{max} = 0.05$ and
$\sin 2\sigma = 1$.
Contour lines of constant $M_\Delta$ are also shown. The efficiency
is expected to be large in the unshaded region. In the shaded regions
closer to the axes (light blue and red), one of the decay channels
$\Delta_s \rightarrow H_u H_u$ or 
$\Delta_s \rightarrow \overline L\, \overline L$
is out of equilibrium. In the larger shaded
region near the $\lambda_L$ axis (light yellow), the decay channel
$\Delta_s \rightarrow \tilde L^c_1 \tilde L^c_1$ is in thermal
equilibrium. 
\Fig{parspace} shows that the observed value of the baryon asymmetry,
$\eta \epsilon_{\Delta} \approx 10^{-8}$, can be achieved in a sizeable portion
of the parameter space. Although the CP asymmetry grows for growing values
of $m_1$, the region in which the decay channel
$\Delta_s \rightarrow \tilde L^c_1 \tilde L^c_1$ is in thermal equilibrium
(implying a small efficiency) becomes simultaneously larger. In the case
of an inverted mass hierarchy, the out-of-equilibrium condition is hardly
satisfied because $|m_{11}|$ is bounded from below.

For completeness, we give an example of parameter choice with
$m_1 \sim m_2$ in the large efficiency region:
$m_1 = 0.005$~eV, $\sin^2 \theta_{13} = 0.05$,
$(\rho, \sigma) = (\pi/4, \pi/2)$, $\lambda_L = 0.1 \gg \lambda_{H_d}$
and $M_\Delta = 10^{12}$ GeV. This gives
$\epsilon_\Delta \simeq 0.8 \times 10^{-6} (M_\Delta / M_{24})$
together with $K_L = 45$, $K_{H_u} = 5.0$ and $K_{L^c_1} = 0.19$
(corresponding to $\lambda_{H_u} = 3.3 \times 10^{-2}$ and
$\lambda_{L^c_1} = 6.5 \times 10^{-3}$). We note in passing that
this choice of parameters corresponds to an effective Majorana mass
$|m_{ee}| = 3.3$ meV for neutrinoless double beta decay.

A comment is in order on the scale of leptogenesis. As can be seen
from \Fig{parspace}, successful leptogenesis in the large efficiency
region requires $M_\Delta \gtrsim 10^{12}$ GeV. Such values
are in strong conflict with the so-called gravitino
constraint, which puts an upper bound
on the reheating temperature after inflation.
From the requirement that the gravitino relic density does not
exceed the dark matter density, one obtains
$T_{RH} \lesssim 10^{(9 - 10)}$ GeV for a gravitino of mass
$m_{3/2} \sim 100$ GeV~\cite{thermal_production}.
If the gravitino is not the LSP, a much stronger bound comes
from the requirement that its decays do not spoil the successful
predictions of Big-Bang Nucleosynthesis~\cite{gravitino_problem},
but it is more model-dependent and can be evaded
in some schemes (e.g. in the presence of $R$-parity
violation~\cite{BCHIY07}). However, there are ways to reconcile
the large efficiency regime of our scenario with the gravitino constraint.
A first possibility is to assume an extremely light gravitino~\cite{PP81},
$m_{3/2} \leq 16$ eV~\cite{VLHMR05}, where the upper
bound comes from WMAP and Lyman-$\alpha$ forest data. 
In this case, the gravitino decouples when it is still relativistic
and escapes the overproduction problem. Such a small
gravitino mass can be obtained in some scenarios of
gauge-mediated supersymmetry breaking~\cite{ITY05}.
A second possibility is to assume a very heavy
gravitino~\cite{Weinberg82},  $m_{3/2} \gg 100$~TeV.
In this case, the gravitino decays
before nucleosynthesis and does not affect the light element
abundances; furthermore, the (neutralino) LSPs produced in the decays
of such heavy gravitinos annihilate efficiently enough to reach
their thermal abundance~\cite{split_Susy}.
Therefore, there is no gravitino problem.
Finally, a third possibility is to resort to non-thermal production
of the heavy triplets, e.g.
during reheating~\cite{reheating}, preheating~\cite{preheating},
or via another mechanism~\cite{other_non_thermal}. In this way
the triplets could be sufficiently produced even though the reheating
temperature lies several orders of magnitude below their mass.

Let us note in passing that the level of predictivity of our leptogenesis
scenario is maintained in a non-supersymmetric version of
the $SO(10)$ model with a real $\bf 54$ Higgs multiplet\footnote{If
the ${\bf 54}$ scalar multiplet were complex, the Lagrangian would
admit two independent couplings of the $\bf 54$ to the ${\bf 10}_i$
fermion multiplets,
${\bf 10}_i {\bf 10}_j \left(f_{ij} {\bf 54} + g_{ij} {\bf 54}^* \right)$,
and the connection between the generated baryon asymmetry
and the light neutrino parameters would be lost.}.
In this case, there is of course no gravitino problem, but the advantages
of supersymmetric unification are lost.

\section{Conclusions \label{sec:conclusions}}

We presented a new leptogenesis scenario in which
the generated baryon asymmetry depends on the low-energy
neutrino parameters. This scenario arises in $SO(10)$ models
with Standard Model fermions split among $\bf 16$ and $\bf 10$
representations and type II realization of the seesaw mechanism.
The predictivity of our scenario is due to the fact that the light neutrino
masses and the CP asymmetry in triplet decays, $\epsilon_\Delta$,
are controlled by the same set of couplings.
This is to be contrasted with most leptogenesis models, in which the
prediction for the CP asymmetry depends on unknown high-energy
flavour parameters.
In our scenario, instead, $\epsilon_\Delta$ is proportional
to $\sum_{k,l} C_{kl}\, \theta (M_\Delta - M_k - M_l)\,
{\rm Im}[m_{kl} (m^*\,m\,m^*)_{kl}]$, where $m$ is the light neutrino
mass matrix, and the coefficients $C_{kl}$ depend on the masses
of the heavy lepton fields, $M_k$ ($k=1,2,3$). The latter are in turn
proportional to the Standard Model charged lepton masses,
up to some degree of
model dependence. As a result, the CP asymmetry in triplet decays
is directly related to the light neutrino parameters; in particular,
the CP violation needed for leptogenesis is provided by the CP-violating
phases of the PMNS matrix.
In the case where $2 M_1 < M_\Delta < M_1 + M_2$, $\epsilon_{\Delta}$
is directly related to the effective Majorana mass of neutrinoless
double beta decay.

We discussed the possibility of generating the observed baryon
asymmetry in two complementary regimes, assuming for definiteness
$M_1 \ll M_\Delta < M_1 + M_2$. The first regime is characterized
by large values of the CP asymmetry and by a strong washout.
As can be seen in \Fig{maximal},
values of the CP asymmetry as large as $10^{-4}$ -- $10^{-3}$ can be
reached for $\theta_{13}$ close to its present upper bound and either
a normal mass hierarchy with $m_1 \approx m_2$, or an
inverted mass hierarchy with $m_3 \approx {\rm few} \times 10^{-2}$ eV.
This allows to generate the observed baryon asymmetry
for efficiencies as small as $10^{-5}$ -- $10^{-4}$. A study of the efficiency
in this regime requires a detailed numerical analysis.
On the contrary, the qualitative features of the large efficiency regime
can be studied analytically. Such a regime takes place in a sizeable
region of the parameter space where
$\lambda_{L^c_1} \ll \lambda_{H_u}, \lambda_L$ and $M_{24} \gg M_\Delta$.
In this region, $\epsilon_{\Delta}$ takes smaller values, as can be seen
in \Fig{parspace}, but still large enough to allow leptogenesis to be successful
provided that $\theta_{13}$ is large and the light neutrinos have a normal
hierarchical mass spectrum.
Upcoming neutrino experiments will further constrain the light neutrino
parameters and make it possible to test the viability of the scenario.

Let us finally comment on other low-energy implications of the
class of $SO(10)$ models in which our leptogenesis scenario takes
place. These models contain heavy states which contribute to the
renormalization of the squark and slepton soft supersymmetry breaking
masses between high and low energies. The pattern of radiative corrections
is not the same as in standard supersymmetric $SO(10)$ models
and may lead to distinctive signatures in flavour and CP violating processes.
Also, the non-standard assignment of matter fields in $SO(10)$
representations may affect the predictions for proton decay.
We defer the exploration of these effects to future work.

\section*{Acknowledgments}
We thank T.~Hambye, H.~Murayama and G.~Senjanovic for useful
discussions. MF, SL and PH were supported in part by the RTN
European Program MRTN-CT-2004-503369 and by
the French Program ÒJeunes Chercheuses et Jeunes ChercheursÓ
of the Agence  Nationale de la Recherche (ANR-05-JCJC-0023).
MF was supported in part by the Marie Curie Intra-European Fellowship 
MEIF-CT-2007-039968. PH was supported in part by
the Marie Curie Excellence Grant MEXT-CT-2004-01429717.



\end{document}